\documentclass[runningheads]{llncs}
\usepackage[T1]{fontenc}
\usepackage{bm} 
\usepackage{graphicx}
\usepackage{mathrsfs}
\usepackage{array}
\usepackage{amsfonts, amssymb}
\usepackage{amsmath}
\usepackage[mathscr]{eucal}
\usepackage{mathtools}
\usepackage{diagbox}
\usepackage{float}
\usepackage{bbding}
\usepackage{xcolor}
\usepackage{hyperref}
\usepackage{ulem}
\usepackage{makecell}
\usepackage{dsfont}
\usepackage{multirow}
\usepackage{cite}
\usepackage{ulem} 
\usepackage{booktabs}

\newcommand{\norm}[1]{\left\lVert#1\right\rVert}

\usepackage{CJKutf8}
\usepackage{changes}

\usepackage{algorithm}
\usepackage{algorithmic}

\makeatletter

\newcommand{\Rmnum}[1]{\expandafter\@slowromancap\romannumeral #1@}
\makeatother

\begin{document}
\title{Multi-modal MRI Translation via Evidential Regression and Distribution Calibration}

\titlerunning{Multi-modal MRI Translation}

\author{Jiyao Liu\inst{1} \and
Shangqi Gao\inst{2}$^{\text{\Envelope}}$  \and
Yuxin Li\inst{1} \and
Lihao Liu\inst{2, 3}  \and 
Xin Gao\inst{1}  \and 
Zhaohu Xing\inst{4}  \and 
Junzhi Ning \inst{3}  \and 
Yanzhou Su \inst{5}  \and 
Xiao-Yong Zhang \inst{6}  \and 
Junjun He \inst{3}  \and 
Ningsheng Xu \inst{1}  \and 
Xiahai Zhuang\inst{1}$^{\text{\Envelope}}$
}

\authorrunning{J Liu et al. }

\institute{Fudan University, Shanghai, China \and
University of Cambridge, Cambridge, United Kingdom \and
Shanghai Artificial Intelligence Laboratory, Shanghai, China \and
The Hong Kong University of Science and Technology (Guangzhou),
Guangzhou, China \and
Fuzhou University, Fujian, China \and
Shanghai Jiao Tong University, Shanghai, China \\
\url{https://zmiclab.github.io/zxh/}}

\maketitle              

\begin{abstract}
Multi-modal Magnetic Resonance Imaging (MRI) translation leverages information from source MRI sequences to generate target modalities, enabling comprehensive diagnosis while overcoming the limitations of acquiring all sequences.
While existing deep-learning-based multi-modal MRI translation methods have shown promising potential, they still face two key challenges: 1) lack of reliable uncertainty quantification for synthesized images, and 2) limited robustness when deployed across different medical centers.
To address these challenges, we propose a novel framework that reformulates multi-modal MRI translation as a multi-modal evidential regression problem with distribution calibration. Our approach incorporates two key components: 1) an evidential regression module that estimates uncertainties from different source modalities and an explicit distribution mixture strategy for transparent multi-modal fusion, and 2) a distribution calibration mechanism that adapts to source-target mapping shifts to ensure consistent performance across different medical centers.
Extensive experiments on three datasets from the BraTS2023 challenge demonstrate that our framework achieves superior performance and robustness across domains.

\end{abstract}

\section{Introduction}

Magnetic resonance imaging (MRI) has become indispensable in modern clinical practice for its ability to provide detailed soft tissue visualization. Multi-modal MRI sequences, such as native T1-weighted (T1n), T2-weighted (T2w), T2-FLAIR (T2f), and post-contrast T1-weighted (T1c), each capture distinct tissue characteristics that together enable comprehensive diagnosis and treatment planning \cite{baid2021rsna}. However, acquiring a complete set of sequences is often impractical due to various constraints including prolonged scanning time, high costs, and patient-specific limitations. Deep learning-based medical image translation \cite{isola2017image, kong2021breaking} has emerged as a promising solution by enabling the synthesis of missing modalities from available sequences. Recent advances in this field have demonstrated remarkable progress by developing multi-modal fusion techniques \cite{liu2023one,dalmaz2022resvit}. Generative adversarial networks \cite{sharma2019missing, zhou2020hi} or diffusion models \cite{xing2024cross, meng2024multi} were employed to produce high-fidelity synthetic images. To achieve multi-modal integration, three key strategies: image-level concatenation \cite{meng2024multi}, feature-level fusion \cite{zhou2020hi,zhang2024unified}, and attention-based fusion mechanisms \cite{han2023explainable},  were developed.

\begin{figure}[!t]
    \centering 
    \includegraphics[width=0.9\textwidth]{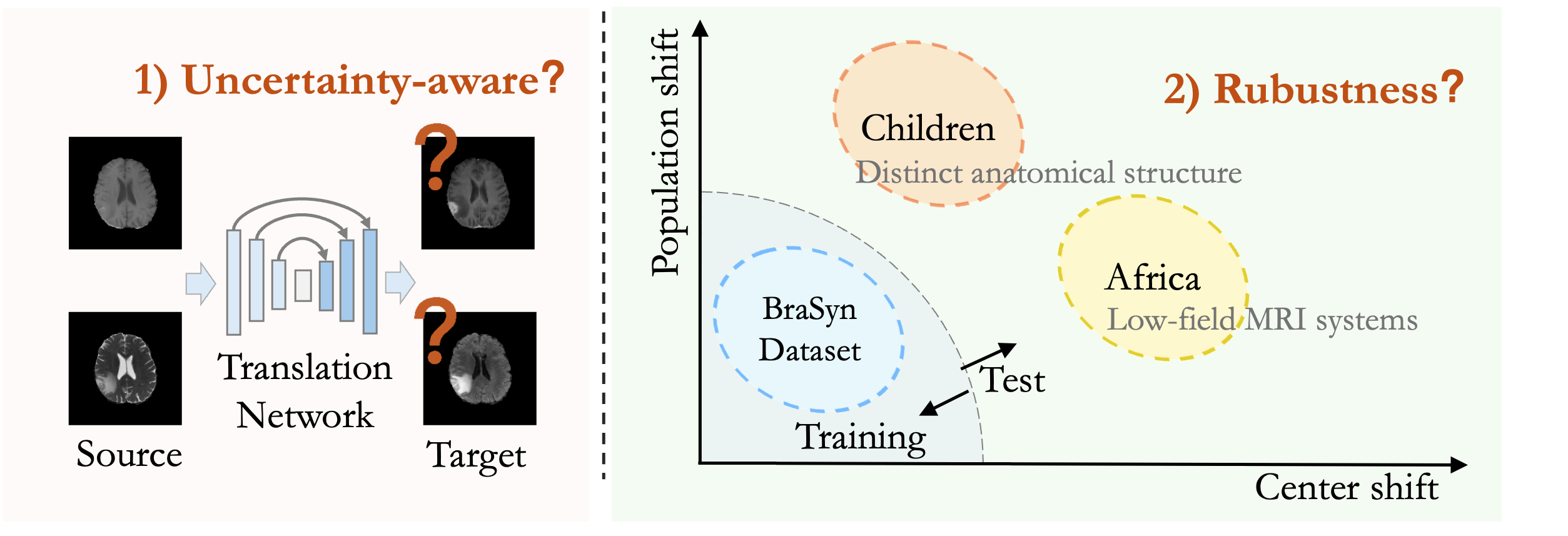}
    
    \caption{Two challenges in multi-modal MRI translation: (1) lack of reliability assessment for synthesized images, and (2) limited robustness across domains.}
    \label{challenge}
    % \vspace{-0.2cm}
\end{figure}

However, despite extensive studies, multi-modal MRI translation still faces two practical challenges (Fig. \ref{challenge}). First, most multi-modal methods lack mechanisms to assess the realiability of synthesized images, making it difficult to determine whether the generated results can be trusted by clinicians. Second, these methods are often sensitive to domain shifts when deployed across centers, where variations in scanning protocols and population characteristics can significantly impact model performance.

To evaluate the reliability of multi-modal learning, few works have focused on the uncertainty-guided fusion strategy \cite{gao2023reliable,han2022trusted,zhang2024biophysics}. Existing uncertainty estimation methods include Bayesian neural networks \cite{maddox2019simple, daxberger2021bayesian} and deep ensembles \cite{buisson2010uncertainty}, which often require complex modifications to network architectures or multiple forward passes. In contrast, evidential learning \cite{sensoy2018evidential, amini2020deep} provides a more efficient and direct approach by learning to predict the parameters of probability distributions. To enhance the reliability of multi-modal fusion, the Mixture of Normal-Inverse Gamma (MoNIG) distribution \cite{ma2021trustworthy} combined predictions from each source through an explicit uncertainty-aware fusion rule. 
To improve model robustness, recent studies have investigated calibration approaches \cite{kuleshov2018accurate,hickey2024transfer} in machine learning. However, their applicability in multi-modal MRI translation is still underexplored.

In this paper, we propose an uncertainty-aware multi-modal MRI translation framework based on evidential regression and distribution calibration. As shown in Fig.~\ref{fig:framework}, we estimate the translation uncertainty of each modality through evidential networks. Based on uncertainties, we apply fusion rules derived from MoNIG to obtain the final generated results, thereby fully leveraging the multi-modal information in a reliable way. Additionally, we improve the robustness across domains through an efficient distribution calibration mechanism based on quantile regression, which enables rapid adaptation to new clinical environments with minimal data requirements.

Our main contributions include three aspects. First, we propose a novel probabilistic framework that reformulates multi-modal MRI translation as an evidential regression problem guided by uncertainty. Second, we develop a distribution calibration mechanism to adapt different domains for multi-modal fusion. Third, extensive experiments on three BraTS2023 datasets demonstrate that our approach delivers superior robustness across different centers.

\section{Methodology}

    \begin{figure*}[t]
        \includegraphics[width=\textwidth]{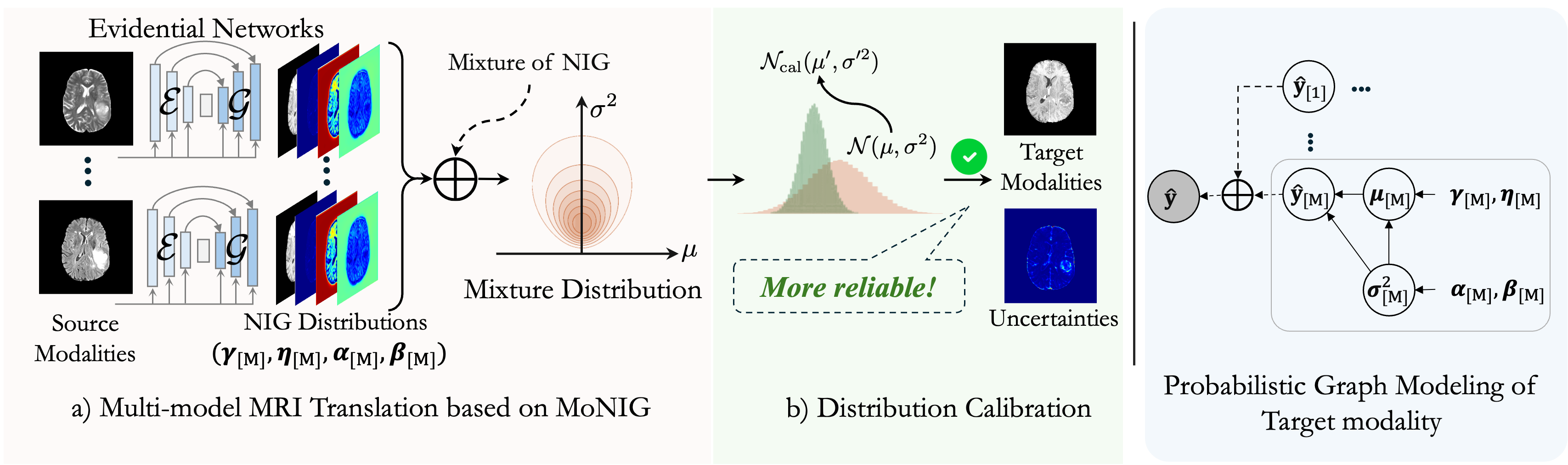}
        \caption{Overview of the proposed framework. Through MoNIG-based image translation, we synthesize a target sequence from existing multi-modal source modalities based on explicit distribution fusion. Then, to adapt to the new source-target mapping, distribution calibration is performed to compute the calibrated generated target images and prediction uncertainty.}
        \label{fig:framework}
        % \vspace{-0.1cm}
    \end{figure*}
    
    The aim of our method is to derive a distribution of target modality with uncertainty and utilize this uncertainty for explicit multi-modal fusion. As shown in Fig. \ref{fig:framework}, our framework mainly consists of two parts, \textit{i.e.}, multi-modal MRI translation with uncertainty quantification and distribution calibration. In the multi-modal MRI translation part, each source modality first estimates the target modality distribution through an evidential network and then obtains a mixture distribution through the explicit MoNIG fusion strategy. Distribution calibration employs quantile regression to obtain the calibrated target distribution, from which the final generated target modality and uncertainty are derived. The details of our method are discussed in the following sections.
    
    \subsection{Uncertainty Estimation and Multi-modal Fusion}
    \subsubsection{Evidential Regression Module}
    \label{subsec:preliminary}
    
    Consider multi-modal medical image translation as a regression problem: given a dataset $\mathcal{D} = \{\{\mathbf x_{[m]}\}_{m=1}^M, \mathbf y\}$, where $\mathbf x_{[m]}$ represents the $m$-th source modality and $\mathbf y$ corresponds to the target modality. Traditional approaches typically train a model by minimizing the $L_1$ or $L_2$ loss between prediction and target. In contrast, we model the synthesized target $\hat{\mathbf y}$ as being drawn from a Gaussian distribution $\hat{\mathbf y} \sim \mathcal{N}(\bm\mu, \bm\sigma^2)$.
    
    To incorporate uncertainty into the model, following \cite{amini2020deep}, we further model the mean and variance using a Normal-Inverse Gamma (NIG) distribution, where $\bm\mu\sim\mathcal{N}(\bm\gamma,\bm\sigma^2\bm\eta^{-1})$ and $\bm\sigma^2\sim\Gamma^{-1}(\bm\alpha,\bm\beta)$. To train this evidential regression model, we minimize the negative log-likelihood loss:
    \begin{equation}
        \mathcal L^{\rm NLL}= \frac{1}{2}\mathrm{log}(\frac{\pi}{\bm\eta}) - \bm\alpha \mathrm{log}(\bm\omega) 
        + (\bm\alpha +\frac{1}{2})\mathrm{log}((\mathbf{y}-\bm\gamma)^2\bm\eta + \bm\omega) + \mathrm{log}(\bm\phi),
    \end{equation}
    where $\bm\omega = 2\bm\beta(1+\bm\eta)$ and $\bm\phi=\frac{\Gamma(\bm\alpha)}{\Gamma(\bm\alpha + \frac{1}{2})}$. We incorporate an additional regularization term:
    \begin{equation}\label{eq:NIGloss}
        \mathcal{L}^{\mathrm{NIG}} = \mathcal{L}^{\mathrm{NLL}} + \lambda_{\mathrm{R}}\mathcal{L}^{\mathrm{R}},
    \end{equation}
    where $\mathcal{L}^{\mathrm{R}} = \norm{\mathbf{y}-\bm\gamma}_1\cdot(2\bm\eta+\bm\alpha)$, and $\lambda_{\mathrm{R}}>0$ balances these two terms.

    Given the trained model, we can compute aleatoric uncertainty ($\textrm{AU}=\frac{\bm\beta}{(\bm\alpha-1)}$) and epistemic uncertainty ($\textrm{EU}=\frac{\bm\beta}{\bm\eta(\bm\alpha-1)}$). After obtaining these NIG distribution parameters, the synthesized target image is derived as $\hat{\mathbf{y}} = \bm\gamma$.

    \subsubsection{Uncertainty-guided Multi-modal Fusion}
    \label{subsec: fusion method}
    
    Given predictions from different source modalities, each characterized by a NIG distribution, they are combined through the NIG summation operation based on MoNIG \cite{ma2021trustworthy}:
    \begin{equation}
        \mathrm{NIG}(\bm\gamma, \bm\eta, \bm\alpha, \bm\beta) = \bigoplus_{m=1}^M \mathrm{NIG}_{[m]}(\bm\gamma_{[m]}, \bm\eta_{[m]}, \bm\alpha_{[m]}, \bm\beta_{[m]}),
    \end{equation}
    where the $\oplus$ operator follows the fusion rule:
    \begin{equation}
        \begin{aligned}
            & \bm\gamma = (\bm\eta_1+\bm\eta_2)^{-1}(\bm\eta_1\bm\gamma_1+\bm\eta_2\bm\gamma_2), \quad \bm\eta = \bm\eta_1 + \bm\eta_2, \quad \bm\alpha = \bm\alpha_1 + \bm\alpha_2 + \frac{1}{2}, \\
            & \bm\beta = \bm\beta_1 + \bm\beta_2 + \frac{1}{2}\bm\eta_1(\bm\gamma_1-\bm\gamma)^2+ \frac{1} {2}\bm\eta_2(\bm\gamma_2-\bm\gamma)^2.
        \end{aligned}
    \end{equation}

    \subsection{Distribution Calibration}
    \label{subsec:distribution_calibration}

    To address the distribution shifts across different medical centers, we propose a non-parametric calibration scheme based on quantile regression \cite{kuleshov2018accurate}. 
    For each target center, we first collect a small pixel-wise calibration dataset $\mathcal{D}$. We use the evidential model to estimate its predictive cumulative distribution function (CDF) $\hat{F}_i(y) = P(\hat{y}_i \leq y)$.  The calibration dataset is then constructed as:
    \begin{equation}
        \label{eq:calibration_data}
        \mathcal{D} = \left\{ \left( F_n(y_n), \hat{P}(F_n(y_n)) \right) \right\}_{n=1}^N,
    \end{equation}
    where $\hat{P}(p) = \frac{|\{y_n | F_n(y_n)\leq p\}|}{N}$ represents the empirical CDF, quantifying the actual coverage proportion of predicted quantiles.

    To ensure monotonicity of the calibrated distribution, we train an isotonic regression model \cite{kuleshov2018accurate} $R: [0,1] \rightarrow [0,1]$ on the calibration dataset to learn the center-specific distribution mapping. The calibrated distribution is then obtained through composition of the learned isotonic regression with the original CDF:
    \begin{equation}
        \tilde{F}(y|x) = R \circ \hat{F}(y|x)
    \end{equation}

    \subsection{Training Paradigm}
    \label{subsec: framework and traning}

    This section shows the network architecture in our framework and explains the training strategy. Our framework consists of a conditional encoder $\mathcal{E}$ and a conditional evidential generator $\mathcal{G}$. For each source modality $\mathbf{x}_{[m]}$, the prediction follows $\mathrm{NIG}_{[m]} = \mathcal{G}(\mathcal{E}(\mathbf{x}_{[m]}))$, where $\mathcal{G}$ outputs four parameter maps $\{\bm\gamma, \bm\eta, \bm\alpha, \bm\beta\}$. We modify an alias-free GAN \cite{song2023alias, karras2021alias} to implement these modules. A softplus activation is applied to ensure positivity constraints on $\{\bm\eta, \bm\alpha, \bm\beta\}$, with $\bm\alpha$ additionally incremented by 1 to satisfy $\bm\alpha>1$.

    The overall training objective combines source-specific losses and fusion loss:
    \begin{equation}
        \mathcal{L} = \sum_{m=1}^M \mathcal{L}_{\mathbf{x}_{[m]}\rightarrow\mathbf{y}} + \mathcal{L}^{\mathrm{Fuse}},
    \end{equation}
    where for each source modality:
    \begin{equation}
        \label{loss_local}
        \mathcal{L}_{\mathbf{x}_{[m]}\rightarrow\mathbf{y}} = \mathcal{L}^{\mathrm{Adv}}_{\mathbf{x}_{[m]}\rightarrow\mathbf{y}} + \lambda_{\mathrm{NIG}}\mathcal{L}^{\mathrm{NIG}}_{\mathbf{x}_{[m]}\rightarrow\mathbf{y}} + \lambda_{\mathrm{Pix}}(\mathcal{L}^{\mathrm{Syn}}_{\mathbf{x}_{[m]}\rightarrow\mathbf{y}} + \mathcal{L}^{\mathrm{Rec}}_{\mathbf{x}_{[m]}\rightarrow\mathbf{x}_{[m]}}),
    \end{equation}
    with $\mathcal{L}^{\mathrm{Syn}}_{\mathbf{x}_{[m]}\rightarrow\mathbf{y}} = \norm{\bm\gamma_{[m]} - \mathbf{y}}_1$ for target synthesis, $\mathcal{L}^{\mathrm{Rec}}_{\mathbf{x}_{[m]}\rightarrow\mathbf{x}_{[m]}} = \norm{\mathcal{G}(\mathbf{x}_{[m]}) - \mathbf{x}_{[m]}}_1$ for self-reconstruction, $\mathcal{L}^{\mathrm{NIG}}_{\mathbf{x}_{[m]}\rightarrow\mathbf{y}}$ for uncertainty estimation as defined in Equ. \ref{eq:NIGloss}, and $\mathcal{L}^{\mathrm{Adv}}_{\mathbf{x}_{[m]}\rightarrow\mathbf{y}}$ for adversarial training to ensure realistic synthesis. The fusion-level loss after MoNIG is defined as:
    \begin{equation}
        \mathcal{L}^{\mathrm{Fuse}} = \mathcal{L}^{\mathrm{Adv}} + \lambda_{\mathrm{NIG}}\mathcal{L}^{\mathrm{NIG}} + \lambda_{\mathrm{Pix}}\mathcal{L}^{\mathrm{Syn}},
    \end{equation}
    where each term serves the same as Equ. \ref{loss_local} but operates on the fused prediction. The balancing weights $\lambda_{\mathrm{NIG}}$ and $\lambda_{\mathrm{Pix}}$ are set to 0.5 and 100 respectively.

\section{Experiments}

    \subsection{Dataset, Evaluation Metrics, and Implementation}

    \subsubsection{Datasets} We conduct experiments using three brain MRI datasets from the BraTS2023 challenge \cite{baid2021rsna,pati2022federated}: BraSyn, BraTS-Africa, and BraTS-PED. Each dataset comprises four registered MRI sequences (T1n, T2w, T2f, and T1c sequences) and expert-annotated tumor segmentation masks.

    \textit{BraSyn dataset (1,251 cases)} serves as our primary training dataset. To evaluate cross-domain robustness, we employ two additional datasets with domain shifts for validation.  \textit{a) BraTS-Africa dataset (60 cases)}  represents variations in imaging protocols, collected from African medical centers using lower-field MRI systems that have lower image contrast and resolution.  \textit{b) BraTS-PED dataset (90 cases)} represents variations in patient populations, focusing on pediatric high-grade glioma of children with distinct anatomical structures due to age-specific brain development.
    
    For data preprocessing, we normalize the intensity values of each modality to $[-1,1]$. The 3D volumes are processed slice-by-slice along the axial plane and resized to $256\times256$. To ensure sufficient foreground information, we discard the first and last 30 slices where the brain regions are typically minimal. For BraSyn dataset, we randomly split the subjects into training, validation, and testing sets with a ratio of 7:1:2.

    \subsubsection{Evaluation Metrics} 
    We employ multiple complementary metrics to comprehensively evaluate our method:
    1) For synthesis quality, we use the peak signal-to-noise ratio (PSNR) and structural similarity index measure (SSIM);
    2) For uncertainty estimation reliability, we calculate the expected uncertainty calibration error (UCE) \cite{laves2020well}, which quantifies the alignment between predicted uncertainties and actual errors through weighted $L_1$-norm;
    3) For clinical utility assessment, we evaluate downstream task performance using the Dice score for multi-modal tumor segmentation. Specifically, we apply a top-performing BraTS 2021 algorithm \cite{luu2021extending} on both synthesized and original sequences.

    \subsubsection{Implementation Details}
    The framework was implemented in PyTorch and trained on a single NVIDIA H100 GPU. We employed the Adam optimizer with an initial learning rate of $2\times10^{-3}$, which was modulated using a cosine annealing warm restart schedule \cite{loshchilov2016sgdr} over $2\times10^5$ iterations. The evidential loss balancing coefficient $\lambda_{\mathrm{R}}$ in Equ. \ref{eq:NIGloss} was gradually increased from 0 to 1 during training to ensure stable convergence. For the MoNIG fusion module, we initialized the concentration parameters $\{\boldsymbol\alpha, \boldsymbol\beta\}$ to small positive values (0.1) to encourage uncertainty-aware learning from the start.

    \subsection{Results}

    \begin{table}[!t]
        \renewcommand\arraystretch{1.2} 
        \centering
        \caption{Comparison of different methods on target sequence synthesis using BraTS2023 Challenge datasets. We conducted experiments using T1n and T2w as input to synthesize T2f and T1c modalities. The results show the average performance across T2f and T1c synthesis.}
        \footnotesize
        \resizebox{0.92\linewidth}{!}{
        \begin{tabular}{p{2.7cm}<{\centering}| p{2.3cm}<{\centering} | p{2.3cm}<{\centering} p{2.3cm}<{\centering} p{2.3cm}<{\centering}}
        \toprule
        Dataset & Method & PSNR$\uparrow$ & SSIM$\uparrow$ & Dice$\uparrow$ \\
        \midrule
        \multirow{7}{*}{\makecell{BraSyn\\(in-dustribution)}} & Pix2pix & 27.12{\scriptsize±4.17} & 0.904{\scriptsize±0.031} & 0.807{\scriptsize±0.102} \\
        & CycleGan & 26.71{\scriptsize±4.31} & 0.903{\scriptsize±0.033} & 0.807{\scriptsize±0.103} \\
        & MM-GAN & 29.06{\scriptsize±3.39} & 0.924{\scriptsize±0.023} & 0.827{\scriptsize±0.096} \\
        & RegGAN & 25.79{\scriptsize±3.95} & 0.888{\scriptsize±0.035} & 0.797{\scriptsize±0.106} \\
        & ResViT & 26.46{\scriptsize±3.33} & 0.908{\scriptsize±0.033} & 0.814{\scriptsize±0.099} \\                
        & M2DN & 31.65{\scriptsize±3.27} & 0.929{\scriptsize±0.029} & \textbf{0.854{\scriptsize±0.093}} \\
        & Ours & \textbf{32.37{\scriptsize±3.09}} & \textbf{0.937{\scriptsize±0.029}} & 0.852{\scriptsize±0.094} \\
        \midrule 
        \multirow{7}{*}{\makecell{BraTS-Africa\\(cross-center)}} & Pix2pix & 20.42{\scriptsize±3.07} & 0.817{\scriptsize±0.170} & 0.749{\scriptsize±0.174} \\
        & CycleGan & 21.16{\scriptsize±2.68} & 0.835{\scriptsize±0.142} & 0.762{\scriptsize±0.165} \\
        & MM-GAN & 21.42{\scriptsize±3.69} & 0.862{\scriptsize±0.165} & 0.782{\scriptsize±0.162} \\
        & RegGAN & 19.22{\scriptsize±2.79} & 0.810{\scriptsize±0.171} & 0.743{\scriptsize±0.178} \\
        & ResViT & 22.12{\scriptsize±3.37} & 0.830{\scriptsize±0.168} & 0.769{\scriptsize±0.168} \\                
        & M2DN & 24.71{\scriptsize±2.29} & 0.858{\scriptsize±0.095} & 0.795{\scriptsize±0.151} \\
        & Ours & \textbf{28.48{\scriptsize±2.17}} & \textbf{0.886{\scriptsize±0.071}} & \textbf{0.828{\scriptsize±0.157}} \\
        \midrule
        \multirow{7}{*}{\makecell{BraTS-PED\\(cross-population)}} & Pix2pix & 22.00{\scriptsize±3.66} & 0.875{\scriptsize±0.045} & 0.667{\scriptsize±0.153} \\
        & CycleGan & 22.32{\scriptsize±3.45} & 0.884{\scriptsize±0.042} & 0.679{\scriptsize±0.144} \\
        & MM-GAN & 22.84{\scriptsize±4.54} & 0.910{\scriptsize±0.037} & 0.700{\scriptsize±0.139} \\
        & RegGAN & 20.91{\scriptsize±3.33} & 0.863{\scriptsize±0.044} & 0.662{\scriptsize±0.156} \\
        & ResViT & 23.83{\scriptsize±3.68} & 0.887{\scriptsize±0.038} & 0.686{\scriptsize±0.146} \\                
        & M2DN & 26.39{\scriptsize±3.05} & 0.870{\scriptsize±0.110} & 0.712{\scriptsize±0.143} \\
        & Ours & \textbf{29.14{\scriptsize±2.92}} & \textbf{0.883{\scriptsize±0.105}} & \textbf{0.723{\scriptsize±0.140}} \\
        \bottomrule
        \end{tabular}}
        \label{compare_tab}
    \end{table}

    \subsubsection{Intra-domain Comparison Study}

    To comprehensively evaluate our framework, we conducted extensive comparisons with other medical image translation methods, including:
    (1) CNN-based methods: Pix2pix \cite{isola2017image}, CycleGAN \cite{zhu2017unpaired}, RegGAN \cite{kong2021breaking}, and MM-GAN \cite{sharma2019missing}; 
    (2) Transformer-based method: ResViT \cite{dalmaz2022resvit}; 
    (3) Diffusion-based method: M2DN \cite{meng2024multi}. 

    As shown in Tab. \ref{compare_tab}, the top section presents the intra-center validation results on the BraSyn dataset. Our method outperforms the second-best approach with improvements of 0.72dB in PSNR and 0.008 in SSIM while achieving competitive Dice scores with the best-performing method. This superior performance is due to our explicit MoNIG fusion strategy, where uncertainty-guided integration is more robust compared to other fusion strategies.

    \subsubsection{Cross-domain Comparison Study}

    To demonstrate robustness, we conducted experiments on two cross-center datasets, BraTS-Africa and BraTS-PED, as shown in Tab. \ref{compare_tab}. To ensure a fair comparison, we randomly selected 10 slices from each cross-center dataset for fine-tuning each comparison model pre-trained on BraSyn as well as training our calibration model.
    
    Our method achieved superior performance on both datasets compared to other approaches, demonstrating the strong robustness of our uncertainty-based framework. Moreover, through calibration training with very few samples, our model could quickly adapt to new source-target mappings.

    \begin{figure}[!t]
        \centering 
        \includegraphics[width=0.9\textwidth]{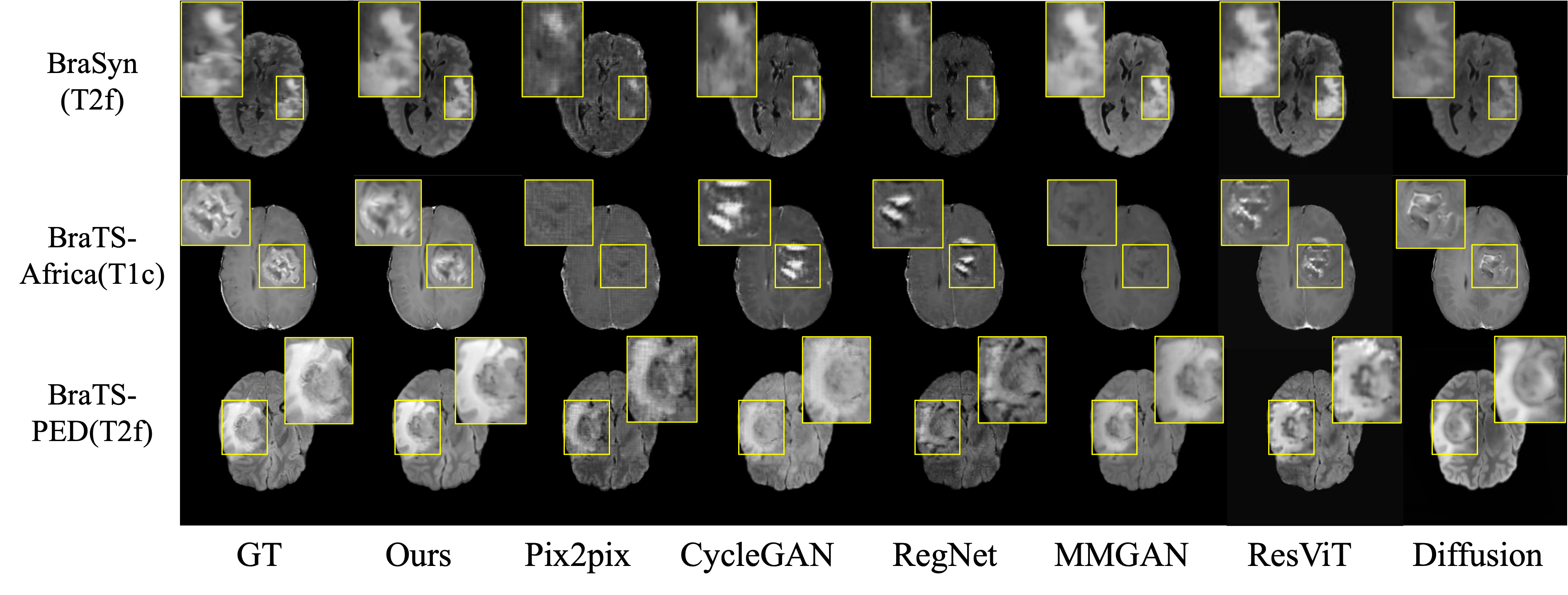}
        
        \caption{Visual comparisons of ours and other methods. }
        \label{compare}
        % \vspace{-0.1cm}
    \end{figure}

    \subsubsection{Ablation Study} 

    \begin{figure}[!t]
        \centering
        \begin{minipage}[t]{0.50\linewidth}
            \centering
            \includegraphics[width=\linewidth]{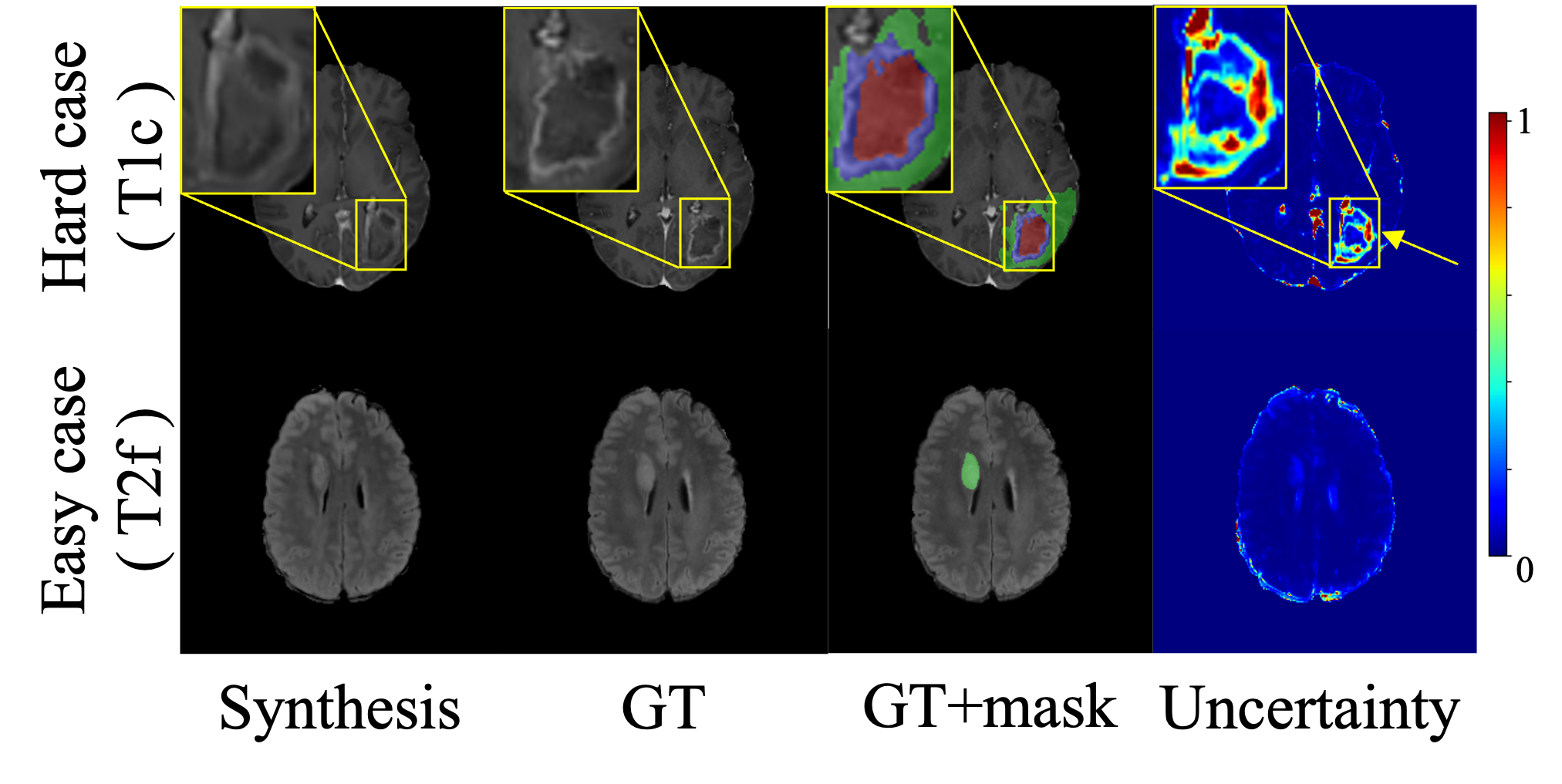}
        \end{minipage}
        % \hfill
        \begin{minipage}[t]{0.45\linewidth}
            \centering
            \renewcommand\arraystretch{1.2}
            \vspace{-3.1cm}
            \resizebox{\linewidth}{!}{
            \begin{tabular}{p{3cm}<{\raggedright} | p{1.3cm}<{\centering} p{1.3cm}<{\centering} p{1.3cm}<{\centering} p{1.3cm}<{\centering}}
                \toprule
                \normalsize Method & \normalsize PSNR$\uparrow$ & \normalsize SSIM$\uparrow$ & \normalsize Dice$\uparrow$ & \normalsize UCE$\downarrow$ \\
                \midrule
                \normalsize Baseline & \normalsize 24.25\newline{\footnotesize±2.42} & \normalsize 0.855\newline{\footnotesize±0.121} & \normalsize 0.783\newline{\footnotesize±0.156} & \normalsize - \\
                \midrule
                \normalsize +Deep Ensemble & \normalsize 25.72\newline{\footnotesize±2.31} & \normalsize 0.864\newline{\footnotesize±0.108} & \normalsize 0.809\newline{\footnotesize±0.152} & \normalsize 0.167\newline{\footnotesize±0.023} \\
                \midrule
                \normalsize +MoNIG & \normalsize 26.39\newline{\footnotesize±2.25} & \normalsize 0.871\newline{\footnotesize±0.095} & \normalsize 0.819\newline{\footnotesize±0.147} & \normalsize 0.142\newline{\footnotesize±0.019} \\
                \normalsize +MoNIG+Calib. & \normalsize \textbf{28.48}\newline\textbf{{\footnotesize±2.17}} & \normalsize \textbf{0.886}\newline\textbf{{\footnotesize±0.071}} & \normalsize \textbf{0.828}\newline\textbf{{\footnotesize±0.157}} & \normalsize \textbf{0.089}\newline\textbf{{\footnotesize±0.012}} \\
                \bottomrule
            \end{tabular}
            }
        \end{minipage}
        \caption{Left: Reliability visualization assessment, where purple mask shows enhanced tumor region, green mask shows oedema region, and red mask shows necrotic core. Right: Ablation study results evaluated on BraTS-Africa dataset.}
        \label{fig:reliability_and_ablation}
        % \vspace{-0.2cm}
    \end{figure}

    We conducted an ablation study to evaluate the effectiveness of our key components. As shown in the right of Fig. \ref{fig:reliability_and_ablation}, MoNIG achieves lower UCE compared to Deep Ensemble  \cite{buisson2010uncertainty}, indicating more accurate uncertainty estimation. Through uncertainty-guided fusion, MoNIG further achieves higher-quality synthesis results. Furthermore, incorporating calibration on top of MoNIG leads to optimal performance in both image synthesis and uncertainty estimation, demonstrating the robustness of our MoNIG and calibration strategies.

    Furthermore, the left of Fig. \ref{fig:reliability_and_ablation} demonstrates the reliability of our uncertainty estimation. We visualize uncertainty maps for hard and easy cases. In the first row (hard case), the yellow arrow indicates a risky region in the generated image, corresponding to the enhancing tumor area. This region shows high uncertainty as it is difficult to identify in T1n and T2w source images, correctly indicating potential errors. In contrast, the second row shows a case with lower uncertainty, suggesting a more reliable synthesis. Such uncertainty visualization can provide valuable guidance for clinical diagnosis by helping physicians identify regions requiring additional verification.

\section{Conclusion}

In this paper, we presented a framework for multi-modal MRI translation that addresses uncertainty quantification and cross-center adaptation. By reformulating the translation task as an evidential regression problem, our approach provides reliable uncertainty estimates that enhance clinical decision-making. The MoNIG fusion strategy effectively integrates information from multiple source modalities. Furthermore, our distribution calibration mechanism enables efficient adaptation to new clinical environments with minimal data requirements.

\newpage

\bibliographystyle{plain} 
\normalem
\bibliography{bib}  

\begin{thebibliography}{10}

\bibitem{amini2020deep}
Alexander Amini, Wilko Schwarting, Ava Soleimany, and Daniela Rus.
\newblock Deep evidential regression.
\newblock {\em Advances in Neural Information Processing Systems}, 33:14927--14937, 2020.

\bibitem{baid2021rsna}
Ujjwal Baid, Satyam Ghodasara, Suyash Mohan, Michel Bilello, Evan Calabrese, Errol Colak, Keyvan Farahani, Jayashree Kalpathy-Cramer, Felipe~C Kitamura, Sarthak Pati, et~al.
\newblock The rsna-asnr-miccai brats 2021 benchmark on brain tumor segmentation and radiogenomic classification.
\newblock {\em arXiv preprint arXiv:2107.02314}, 2021.

\bibitem{buisson2010uncertainty}
La{\"e}titia Buisson, Wilfried Thuiller, Nicolas Casajus, Sovan Lek, and Gael Grenouillet.
\newblock Uncertainty in ensemble forecasting of species distribution.
\newblock {\em Global Change Biology}, 16(4):1145--1157, 2010.

\bibitem{dalmaz2022resvit}
Onat Dalmaz, Mahmut Yurt, and Tolga {\c{C}}ukur.
\newblock Resvit: Residual vision transformers for multimodal medical image synthesis.
\newblock {\em IEEE Transactions on Medical Imaging}, 41(10):2598--2614, 2022.

\bibitem{daxberger2021bayesian}
Erik Daxberger, Eric Nalisnick, James~U Allingham, Javier Antor{\'a}n, and Jos{\'e}~Miguel Hern{\'a}ndez-Lobato.
\newblock Bayesian deep learning via subnetwork inference.
\newblock In {\em International Conference on Machine Learning}, pages 2510--2521. PMLR, 2021.

\bibitem{gao2023reliable}
Zheyao Gao, Yuanye Liu, Fuping Wu, Nannan Shi, Yuxin Shi, and Xiahai Zhuang.
\newblock A reliable and interpretable framework of multi-view learning for liver fibrosis staging.
\newblock In {\em International Conference on Medical Image Computing and Computer-Assisted Intervention}, pages 178--188. Springer, 2023.

\bibitem{han2023explainable}
Luyi Han, Tianyu Zhang, Yunzhi Huang, Haoran Dou, Xin Wang, Yuan Gao, Chunyao Lu, Tao Tan, and Ritse Mann.
\newblock An explainable deep framework: Towards task-specific fusion for multi-to-one mri synthesis.
\newblock In {\em International Conference on Medical Image Computing and Computer-Assisted Intervention}, pages 45--55. Springer, 2023.

\bibitem{han2022trusted}
Zongbo Han, Changqing Zhang, Huazhu Fu, and Joey~Tianyi Zhou.
\newblock Trusted multi-view classification with dynamic evidential fusion.
\newblock {\em IEEE transactions on pattern analysis and machine intelligence}, 45(2):2551--2566, 2022.

\bibitem{hickey2024transfer}
Jimmy Hickey, Jonathan~P Williams, and Emily~C Hector.
\newblock Transfer learning with uncertainty quantification: Random effect calibration of source to target (recast).
\newblock {\em Journal of Machine Learning Research}, 25(338):1--40, 2024.

\bibitem{isola2017image}
Phillip Isola, Jun-Yan Zhu, Tinghui Zhou, and Alexei~A Efros.
\newblock Image-to-image translation with conditional adversarial networks.
\newblock In {\em Proceedings of the IEEE conference on computer vision and pattern recognition}, pages 1125--1134, 2017.

\bibitem{karras2021alias}
Tero Karras, Miika Aittala, Samuli Laine, Erik H{\"a}rk{\"o}nen, Janne Hellsten, Jaakko Lehtinen, and Timo Aila.
\newblock Alias-free generative adversarial networks.
\newblock {\em Advances in Neural Information Processing Systems}, 34:852--863, 2021.

\bibitem{kong2021breaking}
Lingke Kong, Chenyu Lian, Detian Huang, Yanle Hu, Qichao Zhou, et~al.
\newblock Breaking the dilemma of medical image-to-image translation.
\newblock {\em Advances in Neural Information Processing Systems}, 34:1964--1978, 2021.

\bibitem{kuleshov2018accurate}
Volodymyr Kuleshov, Nathan Fenner, and Stefano Ermon.
\newblock Accurate uncertainties for deep learning using calibrated regression.
\newblock In {\em International conference on machine learning}, pages 2796--2804. PMLR, 2018.

\bibitem{laves2020well}
Max-Heinrich Laves, Sontje Ihler, Jacob~F Fast, L{\"u}der~A Kahrs, and Tobias Ortmaier.
\newblock Well-calibrated regression uncertainty in medical imaging with deep learning.
\newblock In {\em Medical Imaging with Deep Learning}, pages 393--412. PMLR, 2020.

\bibitem{liu2023one}
Jiang Liu, Srivathsa Pasumarthi, Ben Duffy, Enhao Gong, Keshav Datta, and Greg Zaharchuk.
\newblock One model to synthesize them all: Multi-contrast multi-scale transformer for missing data imputation.
\newblock {\em IEEE transactions on medical imaging}, 42(9):2577--2591, 2023.

\bibitem{loshchilov2016sgdr}
Ilya Loshchilov and Frank Hutter.
\newblock Sgdr: Stochastic gradient descent with warm restarts.
\newblock {\em arXiv preprint arXiv:1608.03983}, 2016.

\bibitem{luu2021extending}
Huan~Minh Luu and Sung-Hong Park.
\newblock Extending nn-unet for brain tumor segmentation.
\newblock In {\em International MICCAI brainlesion workshop}, pages 173--186. Springer, 2021.

\bibitem{ma2021trustworthy}
Huan Ma, Zongbo Han, Changqing Zhang, Huazhu Fu, Joey~Tianyi Zhou, and Qinghua Hu.
\newblock Trustworthy multimodal regression with mixture of normal-inverse gamma distributions.
\newblock {\em Advances in Neural Information Processing Systems}, 34:6881--6893, 2021.

\bibitem{maddox2019simple}
Wesley~J Maddox, Pavel Izmailov, Timur Garipov, Dmitry~P Vetrov, and Andrew~Gordon Wilson.
\newblock A simple baseline for bayesian uncertainty in deep learning.
\newblock {\em Advances in neural information processing systems}, 32, 2019.

\bibitem{meng2024multi}
Xiangxi Meng, Kaicong Sun, Jun Xu, Xuming He, and Dinggang Shen.
\newblock Multi-modal modality-masked diffusion network for brain mri synthesis with random modality missing.
\newblock {\em IEEE Transactions on Medical Imaging}, 2024.

\bibitem{pati2022federated}
Sarthak Pati, Ujjwal Baid, Brandon Edwards, Micah Sheller, Shih-Han Wang, G~Anthony Reina, Patrick Foley, Alexey Gruzdev, Deepthi Karkada, Christos Davatzikos, et~al.
\newblock Federated learning enables big data for rare cancer boundary detection.
\newblock {\em Nature communications}, 13(1):7346, 2022.

\bibitem{sensoy2018evidential}
Murat Sensoy, Lance Kaplan, and Melih Kandemir.
\newblock Evidential deep learning to quantify classification uncertainty.
\newblock {\em Advances in neural information processing systems}, 31, 2018.

\bibitem{sharma2019missing}
Anmol Sharma and Ghassan Hamarneh.
\newblock Missing mri pulse sequence synthesis using multi-modal generative adversarial network.
\newblock {\em IEEE transactions on medical imaging}, 39(4):1170--1183, 2019.

\bibitem{song2023alias}
Zhiyun Song, Xin Wang, Xiangyu Zhao, Sheng Wang, Zhenrong Shen, Zixu Zhuang, Mengjun Liu, Qian Wang, and Lichi Zhang.
\newblock Alias-free co-modulated network for cross-modality synthesis and super-resolution of mr images.
\newblock In {\em International Conference on Medical Image Computing and Computer-Assisted Intervention}, pages 66--76. Springer, 2023.

\bibitem{xing2024cross}
Zhaohu Xing, Sicheng Yang, Sixiang Chen, Tian Ye, Yijun Yang, Jing Qin, and Lei Zhu.
\newblock Cross-conditioned diffusion model for medical image to image translation.
\newblock In {\em International Conference on Medical Image Computing and Computer-Assisted Intervention}, pages 201--211. Springer, 2024.

\bibitem{zhang2024biophysics}
Lipei Zhang, Yanqi Cheng, Lihao Liu, Carola-Bibiane Sch{\"o}nlieb, and Angelica~I Aviles-Rivero.
\newblock Biophysics informed pathological regularisation for brain tumour segmentation.
\newblock In {\em International Conference on Medical Image Computing and Computer-Assisted Intervention}, pages 3--13. Springer, 2024.

\bibitem{zhang2024unified}
Yue Zhang, Chengtao Peng, Qiuli Wang, Dan Song, Kaiyan Li, and S~Kevin Zhou.
\newblock Unified multi-modal image synthesis for missing modality imputation.
\newblock {\em IEEE Transactions on Medical Imaging}, 2024.

\bibitem{zhou2020hi}
Tao Zhou, Huazhu Fu, Geng Chen, Jianbing Shen, and Ling Shao.
\newblock Hi-net: hybrid-fusion network for multi-modal mr image synthesis.
\newblock {\em IEEE transactions on medical imaging}, 39(9):2772--2781, 2020.

\bibitem{zhu2017unpaired}
Jun-Yan Zhu, Taesung Park, Phillip Isola, and Alexei~A Efros.
\newblock Unpaired image-to-image translation using cycle-consistent adversarial networks.
\newblock In {\em Proceedings of the IEEE international conference on computer vision}, pages 2223--2232, 2017.

\end{thebibliography}

\end{document}